
%
%
\magnification=\magstep1
\hoffset=0.1truecm
\voffset=0.1truecm
\vsize=23.0truecm
\hsize=16.25truecm
\parskip=0.2truecm
\def\newpage{\vfill\eject}

\def\pp{\parshape 2 0.0truecm 16.25truecm 2truecm 14.25truecm}

\def\fun#1#2{\lower3.6pt\vbox{\baselineskip0pt\lineskip.9pt
  \ialign{$\mathsurround=0pt#1\hfil##\hfil$\crcr#2\crcr\sim\crcr}}}
\def\map{\sigma}
\def\ds{ {\Delta \sigma} }

\def\nmin{ {N_{\rm min} } }

\def\reduce{\Sigma}

\def\diam{ {\cal D} }

\def\mapdown#1{\Bigg\downarrow \rlap{ $\vcenter{\hbox{$#1$}}$ } }

\def\norm{ {1 \over \langle \Sigma \rangle} }

\def\x2{ {\bf x}_2 }

%
%
\centerline{}

\bigskip
\bigskip
\centerline{\bf A QUANTITATIVE ANALYSIS OF}
\centerline{\bf IRAS MAPS OF MOLECULAR CLOUDS}
\vskip 0.25in
\centerline{\bf Jennifer J. Wiseman$^1$ and Fred C. Adams$^2$}
\vskip 0.15in
\centerline{\it $^1$Harvard-Smithsonian Center for Astrophysics}
\centerline{\it 60 Garden Street, Cambridge, MA 02138}
\vskip 0.15in
\centerline{\it $^2$Physics Department, University of Michigan}
\centerline{\it Ann Arbor, MI 48109}
\vskip 0.4in

\centerline{\it submitted to the Astrophysical Journal: 17 September 1993}
\centerline{\it revised: 1 April 1994}

\vskip 0.4in

\centerline {\bf Abstract}

We present an analysis of {\it IRAS} maps of five molecular clouds:
Orion, Ophiuchus, Perseus, Taurus, and Lupus.  For the classification
and description of these astrophysical maps, we use a newly developed
technique which considers all maps of a given type to be elements of a
pseudometric space.  For each physical characteristic of interest,
this formal system assigns a distance function (a pseudometric) to the
space of all maps; this procedure allows us to measure quantitatively
the difference between any two maps and to order the space of all
maps.  We thus obtain a quantitative classification scheme for
molecular clouds. In this present study we use the {\it IRAS}
continuum maps at 100$\mu$m and 60$\mu$m to produce column density (or
optical depth) maps for the five molecular cloud regions given above.
For this sample of clouds, we compute the ``output'' functions which
measure the distribution of density, the distribution of topological
components, the self-gravity, and the filamentary nature of the
clouds.  The results of this work provide a quantitative description
of the structure in these molecular cloud regions.  We then order the
clouds according to the overall environmental ``complexity'' of these
star forming regions.  Finally, we compare our results with the
observed populations of young stellar objects in these clouds and
discuss the possible environmental effects on the star formation
process.  Our results are consistent with the recently stated
conjecture that more massive stars tend to form in more ``complex''
environments.

\bigskip
\noindent
{\it Subject headings:} interstellar: molecules -- stars: formation
-- methods: analytical -- methods: data analysis

\newpage
\bigskip
\centerline{\bf 1. INTRODUCTION}
\medskip

Molecular clouds are important constituents of the galaxy
and comprise a substantial fraction of the galactic mass.
The study of molecular clouds is important on two conceptually
different levels:
Molecular clouds can be considered as astrophysical objects in
their own right and therefore studied as evolving astrophysical
systems.  On the other hand, these clouds can be considered as
providing the initial conditions and background environment for
the star formation process.  This present study has two coupled
goals:  (1) We want to provide a classification scheme for
molecular clouds. (2) We want to begin a quantitative study of
the environmental effects on star formation.  In a companion
paper (Adams \& Wiseman 1994; hereafter AW), we extend the
existing formal techniques for the analysis and classification
of astrophysical maps in general.

Although molecular clouds are extremely well studied objects
and a wealth of observational data exists, quantitative
analyses of these clouds are just now being done.  For example,
many authors have tried to ascertain the degree to which
molecular clouds are fractals (e.g., Bazell \& D\'esert 1988;
Dickman, Horvath, \& Margulis 1990; Falgarone, Phillips,
\& Walker 1991). Veeraraghavan \& Fuller (1991) have applied
techniques from the study of large scale structure in the universe
(see Gott, Melott, \& Dickinson 1986)
to the study of molecular clouds.  Houlahan \& Scalo (1992)
have studied molecular cloud structure using a ``tree''
algorithm which represents a large molecular cloud map
as a ``data tree'' (see also Scalo 1990); this method
explicitly searches for hierarchical structure in the clouds.
Stenholm (1990) has studied molecular clouds by using a
statistical analysis of the properties of molecular
line profiles in the maps.

Many studies have focused on defining ``clumps'' in molecular
clouds and finding the clump mass spectrum (see, e.g., Bally
et al. 1987; Carr 1987; Loren 1989; Lada, Bally, \& Stark 1991;
Stutzki \& Gusten 1990; and the review of Blitz 1993).
Langer, Wilson, \& Anderson (1993) have recently completed
a wavelet analysis of interstellar clouds and have found a clump
mass spectrum of the form $dN/dM \sim$ $M^{-5/3}$ for the cloud
Barnard 5 (they also find evidence for hierarchical structure).
Using a different procedure, Williams, de Geus, \& Blitz (1994)
have found the clump mass spectrum for both the Maddalena molecular
cloud (which has an abnormally low level of star formation compared
to other molecular clouds) and the Rosette molecular cloud (a more
typical star forming cloud).  Both clouds have similar cloud mass
spectra $dN/dM \sim$ $M^{-p}$, where $p=1.32$ for the Rosette cloud
and $p=1.44$ for the Maddalena cloud. To summarize, molecular clouds
seem to exhibit clump mass spectra of the form $dN/dM \sim M^{-p}$,
where $p \approx 3/2$ is a rather robust value.

Another useful approach to describing molecular cloud structure
is to consider the cloud's overall ``complexity''.
Myers (1991) has given a qualitative description of several
nearby molecular cloud complexes; he suggests that these star
forming regions can be ordered according to their ``complexity''
and that the environment can affect the nature of the stars
forming within these clouds.  Recently, Wood, Myers, \& Daugherty
(1994; hereafter WMD) have used this same approach to study
{\it IRAS} maps of molecular clouds. Our present analysis is
both a complement and an extension of the work of Myers (1991)
and WMD; we provide a quantitative description and determination of
``complexity'' and other relevant environmental factors which may
influence the star formation process.

In order to provide a classification scheme for molecular clouds,
we must be able to measure the difference between any two clouds
and to order a set of clouds in some physically meaningful manner.
Unfortunately, the science of form description
for complex irregular entities (such as the molecular clouds
in this study) remains poorly developed (Lord \& Wilson 1984).
In this paper, we utilize a new formalism (Adams 1992; hereafter
Paper I) which considers the difference between any two clouds
to be the ``distance'' between two elements of a metric space
(here, the space of all molecular clouds). We then proceed
by constructing distance functions (here, pseudometrics) for
the space (see our companion paper AW and Paper I for further
details regarding this formalism).   As we illustrate in this
paper, this method of form description provides an effective
means of classifying molecular cloud maps.

The results of this present analysis of molecular clouds
allows us (in principle) to test speculations concerning
environmental effects on the star formation process (see,
e.g., Shu, Adams, \& Lizano 1987 for a general review; see
also Lada \& Shu 1990).  For example, one important unsettled issue
is the question of bimodal star formation, i.e., the assertion that
high-mass stars form in different environments than low-mass stars
(see, e.g., Herbig 1962; Mezger \& Smith 1977; see also Zinnecker,
McCaughrean, \& Wilking 1993 for a recent review).  In order to
address this question, we must first be able to quantitatively
describe ``different environments'' for star formation.  The
results of this paper provide such a description.  However, the
other half of the problem -- a description of the populations
of young stellar objects in the clouds -- remains poorly determined
(largely due to observational selection).
As another example, the current
theory of star formation considers molecular cloud cores (the
actual sites of star formation) to be isolated and nearly spherical
in shape.  This theoretical idealization of a star forming site
has been remarkably successful in predicting many properties of
protostellar objects (see, e.g., the review of Shu et al. 1987)
and has a well-defined and calculable signature in the method of
form description used here.  We can thus test how well actual star
forming regions fit this theoretical idealization (see \S 4).

This paper is organized as follows. We define the observational
sample of molecular clouds in \S 2. In \S 3, we review the
formalism for measuring the distance between molecular clouds
and ordering the space of all clouds.  In \S 4 we calculate output
functions for the clouds in our sample and use the results to
determine coordinates for the clouds.  Using these results, we
order the set of clouds according to each physical
characteristic of interest. In \S 5, we discuss the observed
populations of young stellar objects in these clouds and
discuss their relationship with the cloud characteristics
as described in this paper. We conclude in \S 6
with a discussion and summary of our results.

\bigskip
\bigskip
\centerline{\bf 2. THE MAPS}
\medskip

For our sample of molecular clouds, we use maps of
column density (or, equivalently, optical depth) constructed
from the {\it IRAS} all-sky survey.  Our sample includes
five ``well-known'' molecular cloud regions: Orion,
Ophiuchus, Taurus, Perseus, and Lupus.  These
particular regions were chosen because they are
relatively nearby and large amounts of supporting data
exist.  In addition, as pointed out by Myers (1991),
these five regions seem to span a wide range of cloud
characteristics (``complexity'').  We note that our cloud
denoted as ``Orion'' is only part of the overall Orion
molecular cloud complex; our region is part of what is
generally known as ``Orion B'', although we refer to
the region simply as ``Orion'' for this paper.

The column density maps were produced from the observed
100$\mu$m and 60 $\mu$m {\it IRAS} data by
WMD for the Orion, Ophiuchus, Perseus, and Lupus
clouds and by Houlahan \& Scalo (1992) for the Taurus cloud.
Since detailed descriptions of the map
producing process are given in the original papers, we
present only a brief summary here:  The maps begin as
flux density matrices, where each pixel of the map is
a square with sides of 1 arcminute (notice that this angular
size is smaller than the actual satellite resolution, which
is estimated to be 2--3 arcminute).  After subtraction of
background emission, the dust temperature of each pixel
is estimated from the observed 60$\mu$m/100$\mu$m
color temperature.  After the temperature dependence
has been removed from the map, we are left with a map
of the 100$\mu$m optical depth $\tau_{100}$. The resulting
column density maps are shown in Figures 1 -- 5.

Our sample of clouds has a fairly large dynamic range, both
in spatial extent (typically 400 $\times$ 400 pixels) and
in column density (e.g., a factor of $\sim$400 in Taurus).
Each of these clouds is identified with a known complex of
molecular material as mapped in the CO molecule. Notice,
however, that the outer boundaries of our clouds are
chosen, by necessity, rather arbitrarily. As a result,
we can only obtain information about molecular cloud
structure on spatial scales smaller than the map size.
As a reference point, the physical length scale associated
with a 400 arcminute map is $\sim$17.5 pc,
for a ``standard'' distance  to the cloud of 150 pc.
\footnote{$^\dagger$}{The distances to the clouds in Lupus, Taurus,
and Ophiuchus are estimated to lie in the range 140 -- 160 pc.
The distance to Perseus is less well determined but is
thought to lie in the range 200 -- 350 pc.  On the other
hand, distance to the Orion cloud is $\sim$400 pc and hence
the quoted range of physical scales will be larger by a factor
of 8/3 for Orion. }
We can thus study cloud structure over almost two decades in
physical scale: $\sim$ 0.1 -- 10 pc.  This smallest size scale
corresponds roughly to the full-width at half maximum (FWHM)
contour levels of ammonia cores (e.g., Myers \& Benson 1983);
this largest size scale samples the ``large scale structure''
of the molecular clouds.

Since we are using continuum data in this analysis, we
avoid the usual problems associated with line emission,
where each particular line is subject to different
excitation effects.  The interpretation of continuum
data is thus somewhat cleaner.  On the other hand, we have
no velocity information in this sample and are therefore
confined to studying structure in two spatial dimensions
(i.e., in the plane of the sky).  As with all maps taken
in the plane of the sky, the cloud maps of this paper are
subject to projection effects.

One potential problem with these column density maps is
that at sufficiently large optical depths (either large
$A_V$ or large values of column density), the correlation
between $\tau_{100}$ and $A_V$ is no longer linear (see
Jarrett, Dickman, \& Herbst 1989).
This effect becomes significant for visual extinctions
$A_V$ $>$ 10 or so.  Of the five clouds considered here,
only the Orion region has a substantial fraction of its
area with $A_V$ larger than this limit.  We therefore
expect that this problem will not greatly affect our
interpretation of the other four clouds.  The correction for
this calibration problem amounts to a (nonlinear) re-scaling of
the maps.  Fortunately, under such scaling transformations, the
results of our metric-space formalism transform in a simple manner
(see Theorems 1 and 2 of AW).

Another potential problem with these column density maps is
the possible effects of temperature gradients and multiple
temperature components along a single line of sight.
These effects have been studied in detail for one particular
molecular cloud region (Barnard 5) by Langer et al. (1989).
They use a procedure which is similar to the one described
above and construct a map of column density from the
60 and 100$\mu$m {\it IRAS} maps.
They find that resulting column density map correlates
very well with the ${}^{13}{\rm CO}$ map of the same region.
However, they find that the normalization of the column
density map is wrong in the sense that the estimate of
the total cloud mass is different by a factor of 15
from that found using the ${}^{13}{\rm CO}$ data (which is
thought to be a good tracer of the mass).  Thus, column
density maps produced using this method can provide a
good tracer of cloud structure but the absolute values
of the resulting maps can be problematic (see Langer
et al. 1989 for further discussion of these issues).

\bigskip
\bigskip
\centerline{\bf 3. THE FORMAL SYSTEM FOR FORM DESCRIPTION}
\medskip

We utilize a formalism which considers each molecular cloud to
be an element of an abstract space $X$, which corresponds to the
space of all possible molecular clouds (see Paper I; see also
Elizalde 1987). For a given physical characteristic of interest,
this formalism assigns a one-dimensional function (denoted here
as an {\it output function}) to each cloud.  The difference
between any two clouds can then be determined by finding
the difference between their corresponding output functions;
this difference, in turn, is measured using a standard distance
function (denoted as $d$) defined on the space of functions
(for further discussion of distance functions and metric spaces,
see, e.g., Copson 1968).  The ordering of a
set of clouds is accomplished by assigning a real number --
a {\it coordinate} -- to each cloud, where the coordinate is
defined to be the distance between the cloud and a well-defined
reference state (or set of states).  This entire procedure can be
depicted schematically as:
$$\matrix{
\, & \, & X = \Bigl\{ \map \, \big| \, \map \, \, {\rm is}
\, \, {\rm a} \, \, {\rm cloud} \Bigr\} & \, & \, \cr
\, & \, & \, & \, & \, \cr
\, & \, & \mapdown {d \circ \chi } & \, & \, \cr
\, & \, & \, & \, & \, \cr
\, & \, & \Bigl( X, d \circ \chi \Bigr) & \, & \, \cr
\, & \, & \mapdown {d \circ \chi |_{\map_0} } & \, & \, \cr
\, & \, & \, \Bigl\{ {\rm coordinate} \Bigr\} \subset
{\bf R}^+ & \, & \, \cr } $$
In the above diagram, we have used the symbol $\chi$ to represent the
assignment of an output function to a given map.  Thus, we begin with
the space $X$ of all clouds and we assign output functions to each
cloud.  The composition $d \circ \chi$ measures the difference
(distance) between clouds by measuring the difference between
their output functions (as we discuss below, we take $d$ to be
the usual $L_2$ norm). Using this distance function, we make the
original space of clouds into a pseudometric space
$(X, d \circ \chi)$.  Finally, we assign coordinates (which are
positive real numbers) to the maps through the operation denoted
as $d \circ \chi \big|_{\map_0}$, which measures the distance from
the map to the nearest reference map $\map_0$ (see \S 3.4).
We invoke this procedure for each output function of interest.
Further details of this procedure are discussed in Paper I (see also AW).
In the following discussion, we describe the output functions that we
use for the description and study of molecular clouds.

\medskip
\centerline{\it 3.1 Distributions of Density and Volume}
\medskip

One approach to characterizing a molecular cloud to ask
how much of the material is at the highest densities.
We can define the fraction of the material at high densities
in two different ways.  We first determine the fraction $m$
of the mass in the cloud (map) at densities higher than a
given reference $\reduce$:
$$m (\map; \reduce) \equiv
{ \int d^n {\bf x} \, \, \map( {\bf x}) \, \,
\Theta \bigl[ \map({\bf x}) - \reduce \bigr] \over
\int d^n {\bf x} \, \, \map ({\bf x}) } ,  \eqno(3.1)$$
where $\Theta$ is a step function
and where the integrals are taken over the (bounded)
domain $D$ of the map.  Notice that, for a given map $\map$, $m$ is
a function of one variable (namely $\reduce$).  We can also define
an analogous function $v(\map; \reduce)$ which measures the fraction
of the volume (area in a 2-dimensional map) greater than the reference
density $\reduce$:
$$v (\map; \reduce) \equiv
{ \int d^n {\bf x} \, \, \Theta \bigl[ \map({\bf x}) - \reduce \bigr]
\over \int d^n {\bf x} \, \, } .  \eqno(3.2)$$
Given these definitions, we can define a distance between two maps
by measuring the difference between their corresponding output
functions (using either $m$ or $v$), i.e., we
define a pseudometrics $d_m$ and $d_v$ through
$$d_m (\map_A, \map_B) = \Biggr[ \norm \int_0^\infty  d \reduce \,
\big| m (\map_A; \reduce) - m (\map_B; \reduce)
\big|^2 \Biggr]^{1/2} \, ,  \eqno(3.3)$$
$$d_v (\map_A, \map_B) = \Biggr[ \norm \int_0^\infty  d \reduce \,
\big| v (\map_A; \reduce) - v (\map_B; \reduce)
\big|^2 \Biggr]^{1/2} . \eqno(3.4)$$

The output functions $m(\sigma; \reduce)$ and $v(\map; \reduce)$
have another useful interpretation.  Let us define ${\cal P}_m$
to be (minus) the derivative of the function $m$ with respect to
the variable $\reduce$, i.e.,
$${\cal P}_m (\map; \reduce) = - {d m \over d \reduce} =
{ \int d^n {\bf x} \, \, \map ({\bf x}) \delta
\bigl[ \map({\bf x}) - \reduce \bigr] \over
\int d^n {\bf x} \, \, \map ({\bf x}) } \, ,  \eqno(3.5)$$
where we have used the fact that the derivative of a step
function $\Theta$ is a delta function $\delta$.
Similarly, we define ${\cal P}_v$ via
$${\cal P}_v (\map; \reduce) = - {d v \over d \reduce} =
{ \int d^n {\bf x} \, \, \delta
\bigl[ \map({\bf x}) - \reduce \bigr] \over
\int d^n {\bf x} \, \, } \, .  \eqno(3.6)$$
The quantity ${\cal P}_v$ is the probability (per unit surface density)
of a point in the map $\map$ having the surface density $\reduce$.
Similarly, the quantity ${\cal P}_m$ is the probability (weighted
by the mass) of a point in the map $\map$
having the surface density $\reduce$.  It is straightforward to show
that these probability functions are properly normalized, i.e.,
$\int {\cal P}_v d\reduce$ = 1 and $\int {\cal P}_m d\reduce$ = 1.
The interpretation of the derivatives of $m$ and $v$ as probability
distributions greatly facilitates our understanding of how these
output functions behave under various transformations (see AW).

\medskip
\centerline{\it 3.2 Distribution of Components}
\medskip

We now consider a diagnostic which can discriminate between different
geometrical distributions of the high density material.  One way to
accomplish this goal is to count the number of pieces of the cloud
(i.e., topological components)
as a function of threshold density $\reduce$ (see AW). We first
define a reduced space according to
$$X^+_\reduce \equiv \Bigl\{ {\bf x} \in D \, \Big| \,
\map ( {\bf x} ) > \reduce \Bigr\} . \eqno(3.7)$$
For a given threshold density, the space
$(X^+_\reduce, d_E)$ has a well defined number
$n(\map; \reduce)$ of topological components (where
$d_E$ is the usual Euclidean metric).  We can then define
a pseudometric $d_n$ on the space $X$ of all maps through
$$d_n (\map_A, \map_B) = \Biggr[ \norm \int_0^\infty  d \reduce \,
\big| n (\map_A; \reduce) - n (\map_B; \reduce)
\big|^2 \Biggr]^{1/2} . \eqno(3.8)$$
In defining this distance function, we have chosen to consider
where the mass {\it is} rather than where it {\it is not}.
In other words, we do not explicitly consider holes or voids
in the mass distribution.

\medskip
\centerline{\it 3.3 Distribution of Filaments}
\medskip

We also require some description which measures the shapes of
the pieces of the cloud. Given the breakup of a cloud into components
(as described above), we can obtain a measure of the degree to which
the components are filamentary (i.e., stringlike).  We begin with the
usual definition of the diameter $\diam$ of a set $A$, i.e.,
$$\diam (A) \equiv {\rm max} \Bigl\{ \big| {\bf x} - {\bf y}
\big| \, \, \Big| \, \, {\bf x} , {\bf y} \in A
\Bigr\} . \eqno(3.9)$$
For a given threshold density, a molecular cloud map breaks up
into components as described in the previous section; each of
these components has a well defined diameter.  We can also
calculate the area ${\cal A}$ of a given component.  Notice
that for a perfectly round (circular) component, the area and
the diameter are related by the obvious relation $\cal A$ =
$\pi \diam^2/4$.  In order to obtain a measure of the
departure of a given component from a circular shape, we
first define a factor ${\cal F}_j$, which is simply the
inverse of the filling factor for a given component, i.e.,
$${\cal F}_j \equiv { \pi \diam_j^2 \over 4 {\cal A}_j } \, ;
\eqno(3.10)$$
we denote the quantity ${\cal F}_j$ as the ``filament index''
of the $jth$ component. We also define an average factor $f$:
$$f (\map; \reduce) = {1 \over n(\map; \reduce) }
\sum_j \, w_j \, {\cal F}_j , \eqno(3.11)$$
where the sum is taken over all of the components and where
$f$ is  explicitly written as a function of threshold density
$\reduce$.  The quantities $w_j$ are weighting values; we
consider both an unweighted version of the filament index
($w_j$ = 1) and a weighted version in which each ${\cal F}_j$
is weighted by the fraction of material in that component (see AW).
A highly filamentary cloud will thus have a very large value of $f$.
On the other hand, our theoretical idealization of molecular
clouds breaking up into nearly spherical cores suggests that
at sufficiently high threshold density, $f$ should be nearly unity.
The pseudometric $d_f$ on the space of clouds then can be written
$$d_f (\map_A, \map_B) = \Biggr[ \norm \int_0^\infty  d \reduce \,
\big| f (\map_A; \reduce) - f (\map_B; \reduce)
\big|^2 \Biggr]^{1/2} \, ,  \eqno(3.12)$$
where $f$ can be either the weighted or unweighted version of
the filament function [3.11].

\medskip
\centerline{\it 3.4 Assigning Coordinates}
\medskip

In this study, we want to order the set of molecular clouds
in a meaningful way.  However, a metric (or pseudometric) by
itself does not provide a means of ordering a space.
In this formal system, we assign ``coordinates'' (which are simply
positive real numbers) to the clouds (maps) by measuring the
distance from a given map $\map$ to a reference map $\map_0$.
In this case, we follow Paper I and use uniform density maps
($\map_0$ = {\it constant}) as reference maps.  We also follow
Paper I in defining the coordinate to be the distance to the
{\it nearest} uniform density map. Specifically, for a given
pseudometric $d_\chi$, we define the coordinate $\eta_\chi$ by
$$\eta_\chi \equiv {\rm min} \Bigl\{ d_\chi (\map, \map_0)
\Big| \map_0 \, \, {\rm is} \, \, {\rm a} \, \, {\rm uniform}
\, \, {\rm density} \, \, {\rm map} \Bigr\} \, . \eqno(3.13)$$
See Paper I for further details on implementing this
minimization procedure.  In any case, the coordinates represent
a measure of how far a given cloud is from a uniform state;
these coordinates thus provide a measure of the ``complexity''
of the cloud.

In addition to the coordinates obtained from the output functions
described above, we also consider the ``self-gravity'' of the cloud
as an additional coordinate $\eta_w$ (see AW).  Since the maps used
in this study are maps of column density (rather than volume density),
the quantity $\eta_w$ is a two-dimensional measure of ``self-gravity''
of the cloud. We cannot measure the true (three-dimensional)
self-gravity because we do not have full three-dimensional
information.

\bigskip
\centerline{\bf 4. QUANTITATIVE RESULTS FOR MOLECULAR CLOUDS}
\medskip

In this section, we use the formal system described above (see also
Paper I and AW) to compute output functions and coordinates for
the molecular cloud regions in our sample.  We thus obtain a
quantitative description of these clouds and an overall
ordering of their complexity.  These results are sufficient
to clearly distinguish the clouds in our sample.

\medskip
\centerline{\it 4.1 Output Functions}
\medskip

We begin with a discussion of the output functions themselves.
The distributions of density $m(\map; \reduce)$ for the clouds
are shown in the upper panels of Figures 6--10; the corresponding
distributions of volume $v(\map; \reduce)$ are shown by dotted
curves in the same figures.
Notice that the distribution of density is always greater than
the distribution of volume.  This result is general and can be
shown rigorously for all maps (see Appendix H of AW).
In the lower panels of Figures 6 -- 10 we show the probability
functions ${\cal P}_m (\reduce)$, which represent the probability
of a map having the surface density $\reduce$ (see equation [3.5]).
In the figures, we actually plot $\Delta \sigma {\cal P}_m (\reduce)$,
where $\Delta \sigma$ is the estimated uncertainty in the map values.
For this paper, we assume that this uncertainaty has a constant value
of $\Delta \sigma = 5 \times 10^{-6}$ (Doug Wood, private communication;
see also WMD).

The distributions shown in Figures 6 -- 10 are generally
fairly smooth; in particular, the output functions do not
jump suddenly at any given density scale.  Notice that a
description of cloud structure as high density ``clumps''
moving through a very diffuse ``interclump medium'' would
produce a very different signature; in the extreme limit,
a cloud of this type would show a nearly bimodal distribution
in density and hence nearly a step function in $m(\map; \reduce)$,
where the step occurs at the threshold of the interclump medium
(a second step would occur at the density of the clumps).
We thus argue that these clouds exhibit structure that is not
adequately described by the simple picture of clumps embedded
in an interclump medium (``baseballs in air'').  Clumps can
still be present, although they must exhibit a range of
densities in order to produce the smooth $m(\reduce)$ profiles
observed in these clouds. \footnote{$^\dagger$}{Another
possibility is that ``hard'' uniform density clumps exist,
but that they are much smaller than the beam size.  In this case,
the observations are essentially counting the number of
clumps per beam and smooth $m(\reduce)$ profiles can result.}

The distributions of topological components $n(\map; \reduce)$
are shown in the upper panels of Figures 11 -- 15 and the
corresponding distributions of filaments $f(\map; \reduce)$
are shown in lower panels of the same figures.  For both of these
distributions, we must determine the smallest number of pixels per
component that we want to consider as indicative of real structure.
Figures 11--15 show the components with 3 or more pixels as the
upper solid curve and the components with 9 or more pixels as the
lower dashed curve.  In the following subsection, we discuss the
errors involved in these distributions and the reasoning behind
these particular choices.  As a reference point, we note that
WMD include all components with 7 or more pixels in their sample
of ``real'' cloud structures.

The distributions of components show a great deal of structure but no
big surprises.  The clouds break up into $\sim$100 separate components
and the number varies rapidly (both up and down) with the value of the
threshold.  These distributions thus provide a quantitative method of
showing the generally accepted description of molecular clouds as
complex, clumpy, and irregular objects (see, e.g., Blitz 1993; de Geus
et al. 1990). The distributions of filaments show that the average
value of the filament index over most of the threshold range is $\sim$
2 to 3.  There exists a weak tendency for the filament index to
decrease with increasing column density.  This trend is expected if
gravitational forces (which are intrinsically spherical) play an
increasingly larger role at smaller spatial scales (higher column
densities).

\medskip
\centerline{\it 4.2 Error Considerations}
\medskip

Before we can use the output functions described above to
draw conclusions about cloud structure, we must show that
the effects of observational uncertainties on our results are
sufficiently small.  Fortunately, as we discuss below, this
formal system allows us to directly address this issue.

As shown in our companion paper AW, the error in the
distribution of density $m(\map; \reduce)$ is given by
$$\Delta m = \Delta \sigma \Big| {dm \over d \reduce} \Big|
\, , \eqno(4.1)$$
where $\Delta \sigma$ is the uncertainty in the original map.
We can calculate this uncertainty $\Delta m$ directly from the
output functions themselves.  For each cloud, the error $\Delta m$
reaches a maximum value as a function of the threshold $\reduce$
(see Figures 6 -- 10).  The maximum values are 0.08, 0.012, 0.03,
0.01, and 0.028 for the maps of Lupus, Taurus, Perseus, Ophiuchus,
and Orion, respectively. These error estimates represent the
{\it maximum} possible deviation of the distributions $m(\map; \reduce)$
from their true values due to the presence of observational
uncertainties in the maps.  We thus conclude that the errors
in the distributions of density are sufficiently well
controlled within this formal system.  Similarly, we can show
that the errors are also well controlled for the distributions
of volume $v(\map; \reduce)$.

We now consider possible errors in the distribution of
components function $n(\map; \reduce)$.  As discussed in
our companion paper AW, spurious components can arise due
to pixels in the map erroneously sticking up above the
threshold level when the true value of the pixel is
below the threshold level.  These erroneous pixels can
thus produce erroneous ``islands'' that will be counted
as components.  To estimate the size of the errors produced
by this effect, we must first estimate the probability that a
given pixel in the map will be erroneously larger than
the threshold level.  The calculation of this probability
in AW (see their Appendix F) assumes that the errors in the
individual pixels are randomly distributed.  However, the maps
used in this paper are produced from {\it IRAS} maps which have a
3 arcminute resolution, and yet the final column density maps are
given with 1 arcminute pixels.  The errors in these maps thus have
some correlation, but the functional form of this correlation
remains unknown.

To make a start on this problem, we first determine the probability
$P_1$ that a pixel in the map will be erroneously larger than
the threshold level for the case of random error distributions.
The analysis of AW shows that this probability $P_1$ is bounded by
$$P_1 < { {\sqrt \pi} \over 4} \, \Delta \sigma
\Big| {dm \over d \reduce} \Big|_{\rm max} \, = \,
{ {\sqrt \pi} \over 4} \, \Delta m
\big|_{\rm max} \, \eqno(4.2)$$
where we have used equation [4.1] in obtaining the
second equality. Using the results for $\Delta m$ given above,
we find that the probability $P_1$ is bounded by
0.035, 0.0053, 0.013, 0.0044, and 0.012 for the maps of
Lupus, Taurus, Perseus, Ophiuchus, and Orion, respectively.
The maps considered here contain a large number of pixels
$(\sim 10^5)$ and thus the number of possible erroneous
pixels is rather large, 440 -- 3500.  Consequently, the number of
possible spurious components containing only a single pixel
is unacceptably large; we therefore remove from consideration
all components which consist of only a single pixel.
The number of possible spurious components containing
two adjacent pixels is $\propto$ $P_1^2$ and is estimated
to be in the range $\sim$4 -- 250 for the maps in our sample.
We also remove two pixel components form consideration.
In this study, we keep only those components with three or more
pixels. The number of spurious components arising with three or
more pixels is $\propto$ $P_1^3$ and is estimated to lie
in the range 0.02 -- 9. Since this estimate is relatively
conservative and since the distributions of components
typically have values $\sim$100, we conclude that the
errors in the output functions $n(\map; \reduce)$ would be
sufficiently well controlled if we consider only those
components with 3 or more pixels and if the errors in the
pixels were randomly distributed.  As a starting point, we
thus plot the distributions of components with 3 or more pixels
as the solid curves in Figures 11 -- 15.

Next, we must consider the possible effects of correlations
in the errors. Since the original beam size of {\it IRAS}
is about 3 arcminutes, the correlation length for errors
in the maps should not exceed about 3 pixels.  Thus, in
order to take into account the possible effects of correlated
errors, we have calculated separately the number of components with a
given number $N_P$ of pixels, for $N_P = 1 - 9$
(note that a component with 9 or more pixels should be larger than
the error correlation length in all directions).
First, we find that the relative number of components with a
small number of pixels (i.e., $N_P = 1,2,3$) is small (as expected).
\footnote{$^\dagger$}
{Notice that although the bound presented in the previous paragraph
allows for hundreds of erroneous pixels, most of these erroneous
pixels do not produce spurious components.}
Next, we consider the distributions of components where we include
only those components with 9 or more pixels; these distributions are
shown as the dashed curves in Figures 11--15.  Notice that the general
shapes of both the distributions of components and distributions of
filaments are roughly the same for $\nmin = 3$ and $\nmin = 9$ as the
minimum number of pixels per component, although the overall number of
components for the $\nmin = 3$ case is obviously larger.  In addition,
as we find in the next subsection, the ordering of the clouds for
both the components coordinate and the filaments coordinate
are independent of $\nmin$, where $\nmin=3-9$ is the minimum number of
pixels per component.  Finally, we note that WMD use $\nmin=7$ as
the minimum number of pixels per component.  All of these
considerations suggest that the errors in the distributions
of components are reasonably well controlled for this case.

For estimating the possible errors in the distribution of filaments
function $f(\map; \reduce)$, we use the results of Appendix G in
our companion paper AW.  The relative uncertainty is approximately
given by
$${\Delta f \over f} \approx {\Delta n \over n }
\Bigl\{ 1 - {f_S \over f} \Bigr\} \, , \eqno(4.3)$$
where $f_S$ is the average filament index of any possible
spurious components.  As discussed in AW, any spurious
components are likely to be composed of relatively few
pixels and are thus likely to produce an average filament
index $f_S$ which is not far from unity.  The term in brackets
in equation [4.3] is thus likely to be less than unity.
As a result, the relative error in the distribution of
filaments function is generally less than that of the
component function and hence well controlled for our
sample of clouds.

The effect of limited spatial resolution in the original
maps leads to loss of information in the output functions.
However, the size of this effect is somewhat difficult
to determine.  In order to consider this issue, we have
degraded the spatial resolution of one of our maps (Taurus)
and studied the resulting changes in the output functions.
This exercise is described in the Appendix.

\medskip
\centerline{\it 4.3 Coordinates}
\medskip

Given the output functions shown in Figures 6 -- 15, we can
calculate coordinates for the maps (see \S 3.4 and Paper I).
The results are shown in Table 1.  Large values of the coordinates
imply that the cloud is far from a uniform (constant surface density)
state; thus, increasing values of the coordinates imply increasing
``complexity'' of the cloud.

Notice that one should compare different clouds by comparing
the relative sizes of a particular type of coordinate.  One should
{\it not} compare different types of coordinates for the same cloud
because no absolute normalization exists for such a comparison
(see the definitions of the coordinates in \S 3.4; see also
Paper I).

For the filament index
coordinate, we have included both the weighted and unweighted
versions of the output function (see \S 3.3).  Since the
weighted and unweighted versions have similar values,
the characteristic shape of the components must not depend
strongly on the component size (mass).

\vskip0.5truein

$$\matrix{
\, & \eta_m & \eta_v & \eta_n & \eta_f & \eta_{f_w} & \eta_w \cr
 ------        & --- & --- & --- & --- & --- & --- \cr
{\rm Lupus}    & 1.1 & 0.8 & 140 & 11 & 15 & 0.07 \cr
{\rm Taurus}   & 3.4 & 2.7 & 780 & 49 & 57 & 0.06 \cr
{\rm Ophiuchus}& 4.7 & 2.5 & 260 & 63 & 61 & 0.23 \cr
{\rm Perseus}  & 2.2 & 1.8 & 92  & 21 & 19 & 0.30 \cr
{\rm Orion}    & 57  & 1.5 & 947 & 312 & 308  & 2.08 \cr }$$

\vskip0.25truein
\noindent
Table 1. The coordinates for each cloud in the space of maps,
based on the distance of the output functions from that of a
uniform density map. The coordinates shown here correspond to
the output functions for distribution of density ($\eta_m$),
distribution of volume ($\eta_v$),
distribution of components ($\eta_n$),
and distribution of filaments (unweighted $\eta_f$,
weighted $\eta_{f_w}$); the self-gravity coordinate
($\eta_w$) is also given.

\vskip0.35truein

We now briefly consider the effects of errors on the coordinates.
In the previous subsection, we showed that the errors
in the output functions are well controlled in this study.
In other words, the difference between any of the true molecular
cloud maps and the corresponding observed map (with errors) is
small when measured with any of pseudometrics in this paper.
Theorem 3 of Paper I shows that the difference between
the true coordinate (as measured from the ``true'' map)
and the actual observed coordinate (see Table 1) must
also be small. The errors in the coordinates are thus
also well controlled in this study.  This property of
the formal system -- that small errors in the output
functions lead to correspondingly small errors in
the coordinates -- arises precisely because the formalism
is built using metrics (see the proof of Theorem 3 in
Paper I for further discussion).

The coordinates given in Table 1 for the distributions of
components and the distributions of filaments have been
calculated using $\nmin = 3$ as the minimum number of pixels
per component (i.e., using the solid curves in Figures
11 -- 15).  If we use $\nmin = 9$ as the minimum number of
pixels per components (lower dashed curves in Figures 11 -- 15),
we obtain the new coordinates shown in Table 2 below:

\vskip0.5truein

$$\matrix{
\, & \eta_n & \eta_f & \eta_{f_w} \cr
 ------        & --- & --- & --- \cr
{\rm Lupus}    & 112 & 12  & 16 \cr
{\rm Taurus}   & 606 & 49  & 56 \cr
{\rm Ophiuchus}& 176 & 62  & 58 \cr
{\rm Perseus}  &  58 & 22  & 19 \cr
{\rm Orion}    & 571 & 72  & 73 \cr }$$

\vskip0.25truein
\noindent
Table 2. The coordinates for each cloud in the space of maps,
based on the distance of the output functions from that of a
uniform density map. The coordinates shown here correspond to
the output functions for distribution of components ($\eta_n$)
and distribution of filaments (unweighted $\eta_f$, weighted
$\eta_{f_w}$), where we have included only those components
with 9 or more pixels.

\vskip0.35truein

As shown by Table 2, the coordinates for the distributions of
components are decreased as we increase the minimum number of
pixels per component (compare with Table 1).  However, the
ordering of the coordinates for the distributions of components
is the same for both the $\nmin = 3$ (Table 1) and $\nmin = 9$
(Table 2) cases.
On the other hand, the coordinates for the distributions
of filaments (both weighted and unweighted) hardly change at all
as we increase the minimum number of pixels per component.  One
exception to this trend is the case of Orion, where a few components
with less than 9 pixels persist up to very high intensity values and
thus contribute significantly to the coordinate. Notice, however,
that the Orion filament index coordinate is still the
largest in the sample.

\medskip
\centerline{\it 4.4 Ordering the Space of Clouds}
\medskip

Given the coordinates, we can now consider the ordering of the
clouds.  We note that, in general, any ordering will depend
on the physical characteristic being considered. In other words,
there is no {\it a priori} reason why the ordering for one
coordinate (say $\eta_m$) should be the same as that of another
(say $\eta_n$).  However, we find that a relatively well-defined overall
ordering for these five clouds does exist.  For almost all coordinates
given in Table 1, the Orion cloud has by far the largest values.
Similarly, the Lupus cloud generally has the smallest values.
In addition, the Perseus cloud has coordinate values which are
larger than those of Lupus, but smaller than those of the other
clouds.  The largest ambiguity occurs in the comparison of Taurus
and Ophiuchus where Ophiuchus is ``greater'' in terms of $\eta_m$
and $\eta_f$, but Taurus is ``greater'' in terms of $\eta_n$.  Thus,
the results of this analysis suggest the overall ordering
$${\rm Lupus} < {\rm Perseus} < {\rm Taurus} \sim {\rm Ophiuchus}
< {\rm Orion} \, . \eqno(4.4)$$
We note that this ordering is almost the same as the ordering
of the total masses of these clouds.  We have thus found
quantitative evidence for the hypothesis that more massive
clouds produce more complicated structures (Myers 1991).

We note that the ordering given by equation [4.4] was
obtained by directly calculating the coordinates from the
maps shown in Figures 1 -- 5.  Since the Orion cloud is farther
away (by roughly a factor of three) than the other clouds,
and since all maps were taken to have roughly the same
{\it angular} size, we could/should rescale the Orion map
by a factor $\beta \sim 1/3$.  As shown in our companion paper
(see Theorem 1 and its Corollary), this rescaling lowers the
coordinates for Orion by a factor of $1/\sqrt{3}$. Even with this
rescaling, however, Orion is still ``greater'' than the other
clouds in terms of $\eta_m$, $\eta_f$, $\eta_{fw}$, and $\eta_w$.
The only difference that arises is that Taurus has a larger
value of $\eta_n$ than the rescaled value for Orion. Part of
this difference could be due to loss of resolution in Orion.
Even under this rescaling, however, the ordering of equation
[4.4] still applies.  We also note that the Orion map has
a larger {\it area} by a factor of $\sim$9; thus, the number
of components {\it per unit area} in Orion is even smaller
by this additional factor (recall that the distribution of
components function measures the total number of components,
not the number of components per unit area).

The discussion of ordering given above shows that Taurus and Ophiuchus
are similar in their coordinates, i.e., the two clouds are
approximately the same distance from the nearest uniform state.
However, these clouds are still far apart in molecular cloud
space (in other words, the clouds are far from identical,
as has been stressed many times in the literature).
To be more precise, if we compare the clouds for any output
function $\chi$ using the distance
$$d_\chi [ \chi ({\rm Taurus}; \reduce),
\chi ({\rm Oph}; \reduce)] , \eqno(4.5)$$
we find that the clouds are quite different.  We thus emphasize that
the condition of two clouds having similar coordinates, $\eta_\chi
(\map_A)$ $\sim$ $\eta_\chi (\map_B)$, is much weaker than the
condition that the distance $d_\chi (\map_A, \map_B)$ between the
clouds is small (see Theorem 3 of Paper I).

\bigskip
\bigskip
\centerline{\bf 5. POPULATIONS OF YOUNG STELLAR OBJECTS}
\medskip

In the previous section we presented a quantitative description
of the five molecular cloud regions in our sample.  In this section,
we compare the star formation properties of these same regions.
Unfortunately, however, we have no comparable formal system to
determine the star formation properties.  We will proceed by
searching the literature for information concerning the populations
of young stellar objects (YSOs) in these five regions.  This section is
thus, by necessity, less rigorous than the previous one.
We note that YSOs can either be deeply embedded infrared sources
(Class I in the scheme of Adams, Lada, \& Shu 1987) or optically
revealed pre-main-sequence stars (Class II or III); we consider
both types of YSOs in the following discussion.

We first consider the spectral types of the pre-main-sequence
stars found in these molecular cloud regions.  We find that
Orion contains many more {\it hot} stars (spectral types
O B A F and G) than the Lupus and Taurus clouds (see, e.g.,
Cohen \& Kuhi 1979; Larson 1982; see also Genzel \& Stutzki
1989 for a comprehensive review of star formation in Orion).
In fact, the Lupus cloud contains
almost exclusively M stars (Krautter \& Keleman 1987)
and thus produces cooler stars than does Taurus.
On the basis these population studies of young stellar objects
in Orion, Taurus, and Lupus, Myers (1991) argues that the
stellar populations obey an ordering of the form
$${\rm Lupus} < {\rm Taurus} \ll {\rm Orion} \, . \eqno(5.1)$$
We note that this ordering of stellar populations,
although somewhat subjective, is consistent with the
quantitative ordering of the molecular clouds found in
the previous section. The combination of equations [4.4] and
[5.1] thus provides quantitative evidence for the hypothesis
that more massive stars form in ``more complicated'' regions
(Myers 1991; see also WMD).

The YSO populations of the Taurus and Ophiuchus clouds have been
studied and compared by many authors (see, e.g., Wilking \& Lada
1983; Wilking, Lada, \& Young 1989; Cohen, Emerson, \& Beichman
1989; Kenyon et al. 1990; Beichman, Boulanger, \& Moshir 1992;
and the reviews of Zinnecker et al. 1993 and Lada, Strom,
\& Myers 1993).   These two clouds are roughly comparable in
total mass and have approximately the same order of complexity
as determined in the previous section (see equation [4.4]).
However, the young stellar populations of these two regions
seem to have important differences in their luminosity
functions.  In Ophiuchus, the luminosity function is
dominated by embedded (Class I) sources and their number
decreases with decreasing luminosity.  In Taurus, the
luminosity function is dominated by pre-main-sequence
(Class II) objects and their number increases with
decreasing luminosity.  Although the observed luminosity
functions are thus measurably different, the implications
for the underlying mass distribution of the YSO populations
are difficult to assess due to various selection effects
(see the original papers cited above). However, we can
tentatively conclude that the YSO populations is these two
clouds obey the ordering
$${\rm Taurus} < {\rm Ophiuchus} \, . \eqno(5.2)$$
One interesting possibility is that the difference between
the YSO populations in these two clouds results from the
core region of Ophiuchus containing more luminous sources.
In fact, the $K$-band luminosity function is reportedly
different (in the sense of containing brighter sources) in
the core region of Ophiuchus than in the outer lying areas
of the cloud
(Greene \& Young 1992; see also Barsony, Schombert, \&
Kis-Halas 1991; Rieke, Ashok, \& Boyle 1989).

For the Perseus molecular cloud complex, a study of the embedded
(Class I) populations of YSOs has been recently done by
Ladd, Lada, \& Myers (1993). This study directly compares the
embedded population of Perseus to that of the Taurus molecular
cloud.  Compared to sources in Taurus, those in Perseus have
slightly higher luminosities and distinctly redder (weighted
toward longer wavelengths) spectral energy distributions.
Both of these characteristics roughly indicate
that Perseus is forming stars of higher mass than is Taurus.
For the populations of embedded sources in these two clouds,
we thus infer that
$${\rm Perseus} \, \, > \, \, {\rm Taurus} \, . \eqno(5.3)$$
This ordering is consistent with recent work (Ladd, Myers,
\& Goodman 1994) which suggests that NH$_3$ cores in Perseus
have larger mean linewidths than the cores in Taurus.
Notice that the ordering [5.3] of the embedded populations is
different than the ordering of the overall complexity of
the clouds themselves (see equation [4.4]). One possible
reason for this discrepancy between the ordering of the
clouds and the ordering of the YSO populations is
resolution effects.  The distance to the Perseus cloud
is uncertain, but is thought to lie in the range
200 -- 350 pc.  If the distance to Perseus is as large
as 350 pc, then the map of Perseus has a lower spatial
resolution than our map of Taurus.  This loss of resolution
could make Perseus appear ``less complex'' in the analysis
of previous section.

We note that selection effects can arise in any sample of
young stellar objects.  In the present case, however, three
(and possibly four) of our clouds are approximately the same
distance ($\sim$150 pc) away and thus the selection effects
should be similar. The Orion cloud is farther away ($\sim$400 pc)
and thus we should not expect to observe the faintest objects in
this cloud. However, Orion shows the greatest diversity of young
stellar objects and the most evidence for massive star formation;
if Orion were closer, these characteristics would be even
more apparent.  We note, however, that comparisons of
both cloud properties and YSO populations are best done
among clouds at the same distance in order to minimize
differences in resolution effects and selection effects.

To summarize this section, we argue that the populations of
young stellar objects in these five molecular cloud regions
can be ordered in a manner which is roughly consistent with
the overall ordering of clouds according to ``complexity''
as determined in the previous section.  The YSO population
of Orion is unambiguously the ``greatest'' and that of Lupus
is clearly the ``smallest''.  The comparison of Taurus, Perseus,
and Ophiuchus is more problematic.  The YSO populations of
Taurus and Ophiuchus are clearly different, but the sense of
the difference is not clear.  The Perseus cloud appears to
have more luminous (and hence more massive) embedded sources
than Taurus, but Taurus is ``more complex'' according to \S 4.

Another factor that enters into the relationship between
the presently observed structural complexity of clouds
and their (current) YSO populations is the {\it history}
of star formation in the cloud.  For example, if the clouds
have experienced previous episodes of high-mass star
formation (Orion and Ophiuchus are likely examples),
then the energetic effects of these stars can leave
behind substantial signatures of ``complexity'',
even if the stars themselves are no longer present.

We stress that the ordering of YSO populations is presently
in an extremely primitive state.  As more observational results
become available, we can eventually characterize the YSO
populations in a given region in terms of a mass function.
These mass functions can then be used in a manner analogous
to the output functions of the previous section.  In particular,
we can measure the difference between a given mass function and
some standard reference state (e.g., the classic initial
mass function of Salpeter 1955).  We thus obtain
``mass function coordinates'' which can be used to order
the set of YSO populations in a rigorous manner.

\bigskip
\bigskip
\centerline{\bf 6. SUMMARY AND DISCUSSION}
\bigskip
\centerline{\it 6.1 Results on Cloud Classification}
\medskip

The first result of our analysis is that this method of form
description -- metric space techniques and output functions --
provides a workable classification procedure for molecular
cloud maps.  For each output function (corresponding to the
measurement of some physical quantity), this formalism assigns
a coordinate $\eta_\chi$ to each cloud.  In this present
study, we use five different coordinates and thus assign a
5-dimensional ``vector''
$${\bf \eta} \equiv
\bigl( \eta_m, \eta_v, \eta_n, \eta_f, \eta_w \bigr) \, \eqno(6.1)$$
to each cloud in our sample. It is clear from examination of
Table 1 that this set of coordinates is sufficient to clearly
distinguish the clouds in this sample.  Thus, this formal system
provides a useful framework to classify molecular clouds (see
also Paper I and AW). As discussed below (\S 6.2), this formal
system also provides a convenient means of studying cloud
properties and structures.

This classification procedure (so far) only includes
coordinates describing the internal structure of the clouds.
In addition to these coordinates, we can add the obvious
additional numbers describing overall clouds properties
such as total cloud mass, mean magnetic field
strength, mean sound speed, rotation rate, etc.
We also note that the particular output functions used here
provide a beginning for cloud classification and are not
meant to be definitive.  Additional output functions can
and should be incorporated into this formal system as our
understanding of cloud structure increases.

We find that this set of clouds can be ordered in a meaningful
manner (see \S 4 and especially equation [4.4]).  Although
the ordering could, in principle, be different for each
output function/coordinate used, we find that the ordering
of the clouds is roughly consistent for all of the output
functions.  In addition, the observed ordering (equation [4.4])
roughly confirms the conjecture (suggested by Myers 1991)
that the overall complexity of cloud structure increases
with the total mass of the complex (see also WMD).

The method of form description used in this work has
several advantages.  The first advantage is that the
observational errors in the output functions (and hence
the errors in any results inferred from them) can be well
controlled (see \S 4.2 and AW).  We have shown that the errors
in the original {\it IRAS} maps of this study do not produce overly
large errors in the output functions we derive from them
\footnote{$^\dagger$}{We note, however, that in the case of
the distribution of components output functions, the possible
(unknown) correlations between errors in adjacent pixels make
the error analysis difficult for these particular maps.}
Another feature of this formal system is that the results
are either invariant under a large class of
transformations of the maps, or the results transform in a
simple manner (see AW).  For the {\it IRAS} maps used in
this study, e.g., the calibration is uncertain at large values
of visual extinction $A_V$ (see \S 2).  Fortunately,
however, different choices of calibration can be incorporated
into our results by a simple rescaling of the output
functions (Theorem 2 of AW).

\bigskip
\centerline{\it 6.2 Results on Cloud Properties}
\medskip

The results of this work show the absence of any preferred
column density scale in the output functions.  This claim
holds for all of the output functions considered.  Thus,
one result of this study is that, in general,
{\it statements about cloud structure must be made as
a function of the density threshold}.

For our sample of clouds, the distributions of density
$m(\map; \reduce)$ exhibit very smooth behavior (see
Figures 6 -- 10);
in particular, the functions do not jump suddenly at any
threshold level.  Thus, we argue that the naive description
of clouds consisting of high density ``clumps'' moving through
a diffuse ``interclump medium'' is insufficient.  In the output
functions for our observed sample of clouds, no density scales
appear to substantiate the existence of a clump density and/or
an interclump density.

The observed output functions also show that molecular
clouds have a rather large dynamic range in column density.
For the clouds in our sample, the ratio of the peak
column density to the minimum observable value (the
estimated error level in the maps) varies widely but
is always quite large.  This ratio is approximately 30, 400,
75, $10^3$, and $10^5$ for Lupus, Taurus, Perseus,
Ophiuchus, and Orion, respectively.  Although no well
developed theory currently exists to describe or
predict structure in molecular clouds,
the observed large dynamic range in column density in
these objects strongly suggests that some highly nonlinear
process must be at work.  Nonlinear wave motions provide
one possible mechanism to produce this structure (see, e.g.,
Elmegreen 1990; Adams and Fatuzzo 1993; Adams, Fatuzzo,
\& Watkins 1994); cloud fragmentation provides a second
possible mechanism (e.g., Larson 1985).

The component output functions show rather complicated and
nonmonotonic behavior. This finding is consistent with the
assertion that the clouds exhibit hierarchical structure,
where clumps break up into smaller subclumps as the threshold
level is raised (see Houlahan \& Scalo 1992).  We note, however,
that the component output function does not retain information
regarding the spatial positions of the components (or clumps).
As a result, our results do not directly test for the existence
of hierarchical structure.  In the future, additional output
functions can be developed to address this issue.
\footnote{$^\ddagger$}{Thus far, the methods of Houlahan \& Scalo
(1992) and those used here are complementary.  The former use
``data tree'' methods which test for the presence of hierarchical
structure but do not allow for the construction of metrics
(or pseudometrics).  As a result, data tree methods cannot
be used to order a set of clouds according to their complexity.}

The filament output functions provide us with a quantitative
measure of how far real molecular clouds are from spherical
for each threshold level.
Recall that our current theoretical idealization of star forming
regions assumes that cloud cores are spherical at high
densities; we would thus expect to find $f \to 1$ at
large threshold values.  However, as shown in Figures 11--15,
the filament index typically has a value in the range
2--3 at large threshold values.  We thus obtain a somewhat
mixed result: the departure of cloud cores from spherical
symmetry occurs at the factor of two level.  We note that
the smallest size size probed by the maps in this sample
is $\sim$ 0.1 pc, roughly the physical size scale of
the FWHM contour of an ammonia core (this size scale is
small enough to affect protostellar collapse).  Notice that
the findings of this paper for the shapes of regions in column
density maps are roughly consistent with previous determinations
of cloud core shapes from various molecular line observations
(see, e.g., Myers \& Benson 1983; Loren 1989; Myers \& Fuller
1992; Myers et al. 1993).

Previous studies of structure in molecular clouds have
determined that the cloud boundaries generally exhibit fractal
structure (e.g., Bazell \& D\'esert 1988; Dickman, Horvath, \&
Margulis 1990; Falgarone, Phillips, \& Walker 1991). One might
naively expect that fractal structures would be highly filamentary
and hence would have a rather large filament index (as defined
here), and yet the filament index for molecular clouds is not
too far from unity.  Taken together, these two results imply
that while the boundaries of the components (islands) are
indeed fractal, most of the interior of
the components is ``far from the boundary'' in the sense that
most interior points are not affected by the fractal nature
of the boundary.  This same state of affairs obtains in the
classic example of a fractal boundary -- the coastline of
Great Britain (Mandelbrot 1977).  While the coastline is both
fractal and infinite, our British colleagues can walk around
freely in the interior without worrying about stepping off the
island and into the sea.

We have also considered the populations of young stellar
objects in these clouds (\S 5).  As a general rule, the
YSO populations obey an ``ordering'' that is consistent
with the ordering of the clouds according to their overall
complexity (as given in \S 4.4).  We have thus obtained
support for the conjecture that more massive stars tend
to form in more complicated star forming regions. We note,
however, that the current data on YSO populations in these
clouds are insufficient to make this claim definitive.
We also note that not all of the data presented here
support this conjecture (see, e.g., equation [5.3]).
In the future, as more observational data become available,
this question must be studied in more detail.  Thus,
the important question of bimodal star formation --
the assertion that high mass stars form in different
environments than low-mass stars -- unfortunately remains
open. However, significant progress has been made: We now
have a quantitative method to describe different star forming
environments.

\bigskip
\centerline{\it 6.3 Future Work}
\medskip

While this present paper provides a preliminary step toward
a quantitative description of molecular cloud structure,
many directions for future work remain.  Larger samples of
molecular clouds should be studied and more output functions
should be developed (see also AW). As mentioned above, we must
also obtain a better understanding of the populations of
young stellar objects in these clouds.

One obvious generalization of this work is to study
molecular emission line maps of these clouds.  Such maps
contain velocity information which is not present in the column
density maps of this sample. Although the interpretation of
the velocity information is not straightforward, a simple
method of procedure does exist: Define some type of
``Molecular Cloud Hubble Law'' which converts the observed
line-center velocities into a line-of-sight spatial distance
coordinate. One is then left with a density map on a
three-dimensional domain and, as described in Paper I,
the same (but generalized) output functions can be used.
We note that velocity information has often been used to
provide a pseudo third dimension for molecular clouds, e.g.,
in finding clumps and clump mass spectra (see the review of
Blitz 1993).

Embedded magnetic fields provide another important component
of molecular cloud physics.  These fields play an important
role in helping to support the clouds against gravitational
collapse (e.g., Mouschovias \& Spitzer 1976; Shu et al. 1987)
and probably also help determine the formation of substructure
in these clouds (e.g., Elmegreen 1990, 1993; Adams \& Fatuzzo 1993;
Fatuzzo \& Adams 1993).  Polarization maps
(which are thought to trace magnetic field structure)
of molecular cloud regions are now available (e.g., Goodman 1990).
A quantitative analysis of the inferred magnetic field structure
should be performed.  In particular, the degree to which the magnetic
fields lines are tangled (or straight) should be determined.

Finally, we note that a full understanding of molecular clouds
must take into account the physical processes which form both
the cloud substructure and the clouds themselves.  Unfortunately,
these processes are not well understood at present (see, e.g.,
the review of Elmegreen 1993; see also Blitz \& Shu 1990).
As theories are developed to explain
the formation of molecular clouds and their substructure, some
method is required to quantitatively describe the resulting
structures (both the theoretically predicted structures and
the observed structures used to test the predictions).
The formal system used here provides a rigorous method to
describe such structures and the results of this paper
demonstrate the efficacy of this approach.

\vskip 0.6truein
\nobreak
\centerline{Acknowledgements}
\medskip

We would like to thank Debbie Daugherty, Pat Houlahan, Phil Myers,
John Scalo, and Doug Wood for providing us with processed versions
of the {\it IRAS} maps and for useful discussions.  We also would
like to thank Peter Barnes, Brian Boonstra, Dick Canary, Marco Fatuzzo,
Alyssa Goodman, Paul Ho, Lee Mundy, Greg Thorson, and Rick Watkins
for helpful comments and discussions.  Finally, we would like to
thank the referee -- Eugene de Geus -- for many useful comments and
criticisms.  This work was supported by  NASA Grant Nos. NAGW--2802
and NAGW--3121, by an NSF Young Investigator Award, and by funds from
the Physics Department at the University of Michigan.

\vskip 0.4in
\bigskip
\centerline{\bf APPENDIX: A RESOLUTION STUDY}
\medskip

In this appendix, we discuss the effects of loss of spatial
resolution on our results.  We choose one of our maps --
that of Taurus -- for this study.  As described in \S 4,
we have already calculated the output functions and the
coordinates for this map.  We now degrade the resolution
of the map by averaging together every four pixels, i.e.,
we convert the original 400 $\times$ 400 map into a
200 $\times$ 200 map.  This averaging provides a rough
approximation to moving the cloud a factor of two farther
away and viewing the structure through the same telescope.
For this new map, which we denote as Taurus(2), we
calculate the output functions and the coordinates
as before.   We then degrade the resolution a second
time by averaging together every four pixels in the
Taurus(2) map.   We are thus left with a 100 $\times$ 100
map which we denote as Taurus(4).  For this map, we
also calculate the output functions and the coordinates.
Finally, we degrade the map yet another time using the
same method and denote the resulting map as Taurus (8).
The results of this procedure are summarized in Table 3,
where we present the coordinates for the four maps of
Taurus at differing resolutions.

$$\matrix{
\, & \eta_m &  \eta_v & \eta_n & \eta_f & \eta_{f_w} \cr
 ------         & --- & --- & --- & --- & --- \cr
{\rm Taurus}    & 3.38 & 2.69 & 780 & 49 & 57  \cr
{\rm Taurus(2)} & 3.36 & 2.69 & 494 & 47 & 54  \cr
{\rm Taurus(4)} & 3.31 & 2.67 & 267 & 37 & 42  \cr
{\rm Taurus(8)} & 3.20 & 2.63 & 133 & 25 & 31  \cr }$$

\vskip0.25truein
\noindent
Table 3. The coordinates for the Taurus map at four
different spatial resolutions.  The coordinates shown here
correspond to the output functions for distribution of density
($\eta_m$), distribution of volume ($\eta_v$), distribution of
components ($\eta_n$), and distribution of filaments (unweighted
$\eta_f$, weighted $\eta_{f_w}$).

\vskip0.35truein

Table 3 shows that the effects of degrading the resolution
are not overly severe.  The mass fraction and volume fraction
coordinates hardly change when we degrade the resolution by
a factor of four.  In fact, we must present the coordinates
with additional significant figures in order to see the
change.  The number of components coordinate changes the most;
it steadily decreases as the resolution is degraded. This
behavior is expected as the telescope beam averages together
different pixels in the map (see also \S 4 where we discuss
the issue that the pixel size of the maps is smaller than
the original beamsize of the {\it IRAS} satellite).
The filament coordinate
does not change substantially when the resolution is
degraded by a factor of two, but it begins to decrease
when the resolution is degraded by a factor of four.
To summarize, loss of resolution tends to make clouds
appear ``simpler'' and the size of this effect is
quantified by the coordinates given in Table 3.

\newpage
\bigskip
\bigskip
\centerline{\bf REFERENCES}
\medskip

\par\pp
Adams, F. C. 1992, {\sl ApJ}, {\bf 387}, 572 (Paper I)

\par\pp
Adams, F. C., \& Fatuzzo, M. 1993, {\sl ApJ}, {\bf 403}, 142

\par\pp
Adams, F. C., Fatuzzo, M., \& Watkins, R. 1994, {\sl ApJ}, in press

\par\pp
Adams, F. C., Lada, C. J., \& Shu, F. H. 1987, {\sl ApJ}, {\bf 312}, 788

\par\pp
Adams, F. C., \& Wiseman, J. J. 1994, submitted to {\sl ApJ} (AW)

\par\pp
Bally, J., Langer, W. D., Stark, A. A., \& Wilson, R. W. 1987,
{\sl ApJ}, {\bf 312}, L45

\par\pp
Barsony, M., Schombert, J. M., \& Kis-Halas, K. 1991,
{\sl ApJ}, {\bf 379}, 221

\par\pp
Bazell, D., \& D\'esert, F. X. 1988, {\sl ApJ}, {\bf 333}, 353

\par\pp
Beichman, C. A., Boulanger, F., \& Moshir, M. 1992,
{\sl ApJ}, {\bf 386}, 248

\par\pp
Blitz, L. 1993, in Protostars and Planets III, ed. E. Levy and
M. S. Mathews (Tucson: University of Arizona Press), p. 125

\par\pp
Blitz, L., \& Shu, F. H. 1980, {\sl ApJ}, {\bf 238}, 148

\par\pp
Carr, J. S. 1987, {\sl ApJ}, {\bf 323}, 170

\par\pp
Cohen, M., Emerson, J. P., \& Beichman, C. A. 1989,
{\sl ApJ}, {\bf 339}, 455

\par\pp
Cohen, M., and Kuhi, L. V. 1979, {\sl ApJ Suppl.}, {\bf 41}, 473

\par\pp
Copson, E. T. 1968, Metric Spaces (London: Cambridge University Press)

\par\pp
Dickman, R. L., Horvath, M. A., \& Margulis, M. 1990,
{\sl ApJ}, {\bf 365}, 586

\par\pp
Elizalde, E. 1987, {\sl J. Math. Phys.}, {\bf 28} (12), 2977

\par\pp
Elmegreen, B. G. 1990, {\sl ApJ}, {\bf 361}, L77

\par\pp
Elmegreen, B. G. 1993, in Protostars and Planets III,
ed. E. Levy and M. S. Mathews (Tucson: University of
Arizona Press), p. 97

\par\pp
Falgarone, E., Phillips, T. G., \& Walker, C. K. 1991,
{\sl ApJ}, {\bf 378}, 186

\par\pp
Fatuzzo, M., \& Adams, F. C. 1993, {\sl ApJ}, in press

\par\pp
Genzel, R., \& Stutzki, J. 1989, {\sl A R A \& A}, {\bf 27}, 41

\par\pp
de Geus, E. J., Bronfman, L., \& Thaddeus, P. 1990,
{\sl A \& A}, {\bf 231}, 137

\par\pp
Goodman, A. A. 1990, PhD Thesis, Harvard University

\par\pp
Gott, J. R., Melott, A. L., \& Dickinson, M. 1986,
{\sl ApJ}, {\bf 306}, 341

\par\pp
Greene, T. P., \& Young, E. T. 1992, {\sl ApJ}, {\bf 395}, 516

\par\pp
Herbig, G. H. 1962, {\sl Adv. Astron. Astrophys.}, {\bf 1}, 47

\par\pp
Houlahan, P., \& Scalo, J. M. 1992, {\sl ApJ}, {\bf 393}, 172

\par\pp
Jarrett, T. H., Dickman, R. L., \& Herbst, W. 1989,
{\sl ApJ}, {\bf 345}, 881

\par\pp
Kenyon, S. J., Hartmann, L. W., Strom, K. M., \& Strom, S. E.
1990, {\sl AJ}, {\bf 99}, 869

\par\pp
Krautter, J., \& Keleman, J. 1987, {\sl Mitt. Astron. Ges.},
{\bf 70}, 397

\par\pp
Lada, C. J., \& Shu, F. H. 1990, {\sl Science}, {\bf 1111}, 1222

\par\pp
Lada, E. A., Strom, K. M., \& Myers, P. C. 1993,
in Protostars and Planets III, ed. E. Levy and M. S. Mathews
(Tucson: University of Arizona Press), p. 245

\par\pp
Ladd, E. F., Lada, E. A., \& Myers, P. C. 1993, {\sl ApJ}, {\bf 410}, 168

\par\pp
Ladd, E. F., Myers, P. C., \& Goodman, A. A. 1994, {\sl ApJ}, in press

\par\pp
Langer, W. D., Wilson, R. W., Goldsmith, P. F.,
\& Beichman, C. A. 1989, {\sl ApJ}, {\bf 337}, 355

\par\pp
Langer, W. D., Wilson, R. W., \& Anderson, C. H. 1993,
{\sl ApJ}, {\bf 408}, L45

\par\pp
Larson, R. B. 1982, {\sl M N R A S}, {\bf 200}, 159

\par\pp
Larson, R. B. 1985, {\sl M N R A S}, {\bf 214}, 379

\par\pp
Lord, E. A., \& Wilson, C. B. 1984, {The Mathematical
Description of Shape and Form} (Sussex: Ellis Horwood)

\par\pp
Loren, R. B. 1989, {\sl ApJ}, {\bf 338}, 902

\par\pp
Mandelbrot, B. 1977, {The Fractal Geometry of Nature}
(New York: Freeman)

\par\pp
Mezger, P. G., \& Smith, L. F. 1977, in {Star Formation}
(IAU Symposium 75), eds. T. de Jong and A. Maeder
(Dordrecht: Reidel), p. 133

\par\pp
Mouschovias, T. Ch., \& Spitzer, L. 1976,
{\sl ApJ}, {\bf 210}, 326

\par\pp
Myers, P. C. 1991, in {Fragmentation of Molecular Clouds
and Star Formation} (IAU Symposium 147), eds. E. Falgarone and
G. Duvert (Dordrecht: Kluwer), in press

\par\pp
Myers, P. C., \& Benson, P. J. 1983, {\sl ApJ}, {\bf 266}, 309

\par\pp
Myers, P. C., \& Fuller, G. A. 1992, {\sl ApJ}, {\bf 396}, 631

\par\pp
Myers, P. C., \& Fuller, G. A., Goodman, A. A., \& Benson, P. J.
1991, {\sl ApJ}, {\bf 376}, 561

\par\pp
Rieke, G. H., Ashok, N. M., \& Boyle, R. P. 1989,
{\sl ApJ}, {\bf 339}, L71

\par\pp
Salpeter, E. E. 1955, {\sl ApJ}, {\bf 121}, 161

\par\pp
Scalo, J. 1990, in {Physical Processes in Fragmentation
and Star Formation}, eds. R. Capuzzo-Dolcetta et al.
(Dordrecht: Kluwer), p. 151

\par\pp
Shu, F. H., Adams, F. C., \& Lizano, S. 1987,
{\sl A R A \& A}, {\bf 25}, 23

\par\pp
Stenholm, L. G. 1990, {\sl A \& A}, {\bf 232}, 495

\par\pp
Stutzki, J., \& Gusten, R. 1990, {\sl ApJ}, {\bf 356}, 513

\par\pp
Veeraraghavan, S., \& Fuller, G. A. 1991, in
{Fragmentation of Molecular Clouds and Star Formation}
(IAU Symposium 147), eds. E. Falgarone and G. Duvert
(Dordrecht: Kluwer), in press

\par\pp
Wilking, B. A., \& Lada, C. J. 1983, {\sl ApJ}, {\bf 274}, 698

\par\pp
Wilking, B. A., Lada, C. J., \& Young, E. T.
1989, {\sl ApJ}, {\bf 340}, 823

\par\pp
Williams, J. P., \& Blitz, L. 1993, {\sl ApJ}, {\bf 405}, L75

\par\pp
Williams, J. P., de Geus, E., \& Blitz, L. 1994,
{\sl ApJ}, in press

\par\pp
Wood, D. O. S., Myers, P. C., \& Daugherty, D. A. 1994,
{\sl ApJ Suppl.}, in press (WMD)

\par\pp
Zinnecker, H., McCaughrean, M. J., \& Wilking, B. A. 1993,
in Protostars and Planets III, ed. E. Levy and M. S. Mathews
(Tucson: Univ. of Arizona Press), p. 429

\newpage
\bigskip
\bigskip
\centerline{\bf FIGURE CAPTIONS}
\medskip

\medskip
\noindent
Figure 1.  Column density map of the Lupus molecular cloud region.
This map was taken from the results of WMD and is centered on
the coordinates R.A. = 15$^{\rm h}$ 40$^{\rm m}$ and DEC =
--35$^\circ$ $00'$ (the map extends 200 minutes of arc in
each direction from the center).

\medskip
\noindent
Figure 2.  Column density map of the Taurus molecular cloud region.
This map was taken from the results of Houlahan \& Scalo (1992)
and is centered on the coordinates R.A. = 04$^{\rm h}$ 30$^{\rm m}$
and DEC = 27$^\circ$ $00'$ (the map extends 200 minutes of arc in
each direction from the center).

\medskip
\noindent
Figure 3.  Column density map of the Perseus molecular cloud region.
This map was taken from the results of WMD
and is centered on the coordinates R.A. = 03$^{\rm h}$ 30$^{\rm m}$
and DEC = 32$^\circ$ $00'$ (the map extends 200 minutes of arc in
each direction from the center).

\medskip
\noindent
Figure 4.  Column density map of the Ophiuchus molecular cloud region.
This map was taken from the results of WMD
and is centered on the coordinates R.A. = 16$^{\rm h}$ 25$^{\rm m}$
and DEC = --24$^\circ$ $00'$ (the map extends 200 minutes of arc in
each direction from the center).

\medskip
\noindent
Figure 5.  Column density map of the Orion molecular cloud region.
This map was taken from the results of WMD
and is centered on the coordinates R.A. = 05$^{\rm h}$ 47$^{\rm m}$
and DEC = 00$^\circ$ $00'$ (the map extends 200 minutes of arc in
each direction from the center).

\medskip
\noindent
Figure 6. Upper panel shows the distribution of density
[$m( \reduce)$, solid curve] and distribution of volume
[$v( \reduce)$, dashed curve] output functions for the
Lupus cloud. Lower panel shows the corresponding
probability distribution $\ds {\cal P}_m (\reduce)$ =
$-(dm/d\reduce) \ds$; this function represents the probability
of a pixel in the map having a value $\reduce$ of
threshold column density.

\medskip
\noindent
Figure 7. Upper panel shows the distribution of density
[$m( \reduce)$, solid curve] and distribution of volume
[$v( \reduce)$, dashed curve] output functions for the
Taurus cloud. Lower panel shows the corresponding
probability distribution $\ds {\cal P}_m (\reduce)$ =
$-(dm/d\reduce) \ds$; this function represents the probability
of a pixel in the map having a value $\reduce$ of
threshold column density.

\medskip
\noindent
Figure 8. Upper panel shows the distribution of density
[$m( \reduce)$, solid curve] and distribution of volume
[$v( \reduce)$, dashed curve] output functions for the
Perseus cloud. Lower panel shows the corresponding
probability distribution $\ds {\cal P}_m (\reduce)$ =
$-(dm/d\reduce) \ds$; this function represents the probability
of a pixel in the map having a value $\reduce$ of
threshold column density.

\medskip
\noindent
Figure 9. Upper panel shows the distribution of density
[$m( \reduce)$, solid curve] and distribution of volume
[$v( \reduce)$, dashed curve] output functions for the
Ophiuchus cloud. Lower panel shows the corresponding
probability distribution $\ds {\cal P}_m (\reduce)$ =
$-(dm/d\reduce) \ds$; this function represents the probability
of a pixel in the map having a value $\reduce$ of
threshold column density.

\newpage
\medskip
\noindent
Figure 10. Upper panel shows the distribution of density
[$m( \reduce)$, solid curve] and distribution of volume
[$v( \reduce)$, dashed curve] output functions for the
Orion region. Lower panel shows the corresponding
probability distribution $\ds {\cal P}_m (\reduce)$ =
$-(dm/d\reduce) \ds$; this function represents the probability
of a pixel in the map having a value $\reduce$ of
threshold column density.

\medskip
\noindent
Figure 11.  Upper panel shows the distribution of components
output function $n( \reduce)$ for the Lupus cloud
(where $\reduce$ is the threshold column density).
Lower panel shows the corresponding distribution of filaments
function $f( \reduce)$.  In both panels, the solid curve shows
the distribution which includes all components with 3 or more
pixels; the dashed curve shows the distribution which includes
only those components with 9 or more pixels.

\medskip
\noindent
Figure 12.  Upper panel shows the distribution of components
output function $n( \reduce)$ for the Taurus cloud
(where $\reduce$ is the threshold column density).
Lower panel shows the corresponding distribution of filaments
function $f( \reduce)$. In both panels, the solid curve shows
the distribution which includes all components with 3 or more
pixels; the dashed curve shows the distribution which includes
only those components with 9 or more pixels.

\medskip
\noindent
Figure 13.  Upper panel shows the distribution of components
output function $n( \reduce)$ for the Perseus cloud
(where $\reduce$ is the threshold column density).
Lower panel shows the corresponding distribution of filaments
function $f( \reduce)$.  In both panels, the solid curve shows
the distribution which includes all components with 3 or more
pixels; the dashed curve shows the distribution which includes
only those components with 9 or more pixels.

\medskip
\noindent
Figure 14.  Upper panel shows the distribution of components
output function $n( \reduce)$ for the Ophiuchus cloud
(where $\reduce$ is the threshold column density).
Lower panel shows the corresponding distribution of filaments
function $f( \reduce)$.  In both panels, the solid curve shows
the distribution which includes all components with 3 or more
pixels; the dashed curve shows the distribution which includes
only those components with 9 or more pixels.

\medskip
\noindent
Figure 15.  Upper panel shows the distribution of components
output function $n( \reduce)$ for the Orion region
(where $\reduce$ is the threshold column density).
Lower panel shows the corresponding distribution of filaments
function $f( \reduce)$.  In both panels, the solid curve shows
the distribution which includes all components with 3 or more
pixels; the dashed curve shows the distribution which includes
only those components with 9 or more pixels.

\bye